\documentclass[ASNA,STIX2COL]{WileyNJD-v2}
\usepackage{graphicx}
\usepackage{natbib}

\newcommand\BibTeX{{\rmfamily B\kern-.05em \textsc{i\kern-.025em b}\kern-.08em
T\kern-.1667em\lower.7ex\hbox{E}\kern-.125emX}}

\articletype{Original paper}%

\newcommand{\mygi}{MyGIsFOS}

\newcommand{\logg}{\ensuremath{\log\,g}}
\def\teff{$T\rm_{eff}$}
\newcommand{\kms}{$\rm km s ^{-1}$}
\newcommand*\farcs{\ensuremath{\overset{\prime\prime}{.}}}
\DeclareRobustCommand{\ion}[2]{\textup{#1\,\textsc{\lowercase{#2}}}}
\newcommand*\aap{A\&A}

\newcommand*\ao{Appl.~Opt.}

\newcommand*\apjl{ApJ}

\newcommand*\apjs{ApJS}

\received{<day> <Month>, <year>}
\revised{<day> <Month>, <year>}
\accepted{<day> <Month>, <year>}

\begin{document}

\title{The Gaia RVS benchmark stars. III. Confirmed and new high radial velocity stars from Gaia\,DR3.
\protect\thanks{Based on observations made with UVES at VLT 110.23V0.001 and 112.25D7.001.}}

\author[1]{Elisabetta Caffau*}

\author[1]{David Katz}

\author[1]{Piercarlo Bonifacio}

\author[1]{Ana G\'{o}mez}

\author[1]{Rosine Lallement}

\author[2]{Paola Sartoretti}

\author[1]{Fr\'{e}d\'{e}ric Royer}

\author[2]{Pasquale Panuzzo}

\author[1]{Monique Spite}

\author[3,4]{Patrick Fran\c{c}ois}

\author[2]{Nicolas Leclerc}

\author[5]{Luca Sbordone}

\author[6]{Fr\'{e}d\'{e}ric Th\'evenin}

\author[7]{Hans-G\"unter Ludwig}

\author[8]{Laurent Chemin}

\authormark{E. Caffau \textsc{et al}}

\address[1]{LIRA, Observatoire de Paris, Universit{\'e} PSL, Sorbonne Universit{\'e}, Universit{\'e} Paris Cité, CY Cergy Paris Universit{\'e}, CNRS, 92190 Meudon, France}

\address[2]{UNIDIA, Observatoire de Paris, Universit{\'e} PSL, CNRS,92190 Meudon, France}

\address[3]{LIRA, Observatoire de Paris, Universit{\'e} PSL, Sorbonne Universit{\'e}, Universit{\'e} Paris Cité, CY Cergy Paris Universit{\'e}, CNRS, 75014 Paris, France}

\address[4]{UPJV, Universit\'e de Picardie Jules Verne,33 rue St Leu, Amiens, 80080, France}

\address[5]{European Southern Observatory, Alonso de Cordova 3107, Vitacura, Santiago, Chile}

\address[6]{Universit\'e C\^ote d’Azur, Observatoire de la C\^ote d’Azur, CNRS, Laboratoire Lagrange, France}

\address[7]{Landessternwarte - Zentrum f\"ur Astronomie der Universit\"at Heidelberg, K\"onigstuhl 12, 69117 Heidelberg, Germany}

\address[8]{ CNRS, Observatoire astronomique de Strasbourg, Universit\'e de Strasbourg, UMR 7550, F-67000 Strasbourg, France}

\corres{*Elisabetta Caffau, \email{Elisabetta.Caffau@obspm.fr}}

\abstract[Abstract]{High-velocity stars are interesting targets to unveil the formation of the Milky Way.
In fact they can be recently accreted from an infalling dwarf galaxies or they can be the result of a turbulent merging of galaxies.
Gaia is providing the community a way to select stars for their kinematics and the radial velocity, one of the speed components, is derived from the Gaia RVS spectrum.
High absolute radial velocity values are sensitive to be false positive. 
They are rare and as such they are more easily impacted than lower velocity stars, by contamination by very low SNR spurious measures.
We here investigate a sample of 26 stars with Gaia absolute radial velocity in excess of 500 km/s with spectroscopic follow-up observations with UVES. 
For all but one star the extreme radial velocity is confirmed and these stars are all metal-poor and, as expected, enhanced in the $\alpha$ elements. The complete chemical inventory confirm the large star-to-star scatter for the heavy elements and their on average larger ratio over iron with respect to the Sun, in agreement with the literature.  
}


\maketitle

\section{Introduction}\label{intro}

The two latest Gaia releases \citep{gaiadr2,gaiadr3} provide us with information on the stellar position and movement 
so that stars can now be selected and tagged by their kinematics.
Also, the combination of chemical and kinematical properties of the stars allows us to obtain a deeper knowledge on the formation and evolution of our Galaxy,
highlighting the major accretion episodes \citep[see e.g.][]{belokurov18}.
In the large Gaia archive, there are stars with extreme kinematics, and the question we are confronted with is always: can we trust it?
These kinematically hot stars are a minority of the Galactic population, but, if their kinematics is trustable, they can be stars escaping from the Galaxy \citep[see e.g. LAMOST J115209.12+120258.0 by][]{boubert2018}, or
falling in it \citep[see e.g.][]{ghs143_2024}, or belonging to a stream,  and they can be there to tell us a story.

For a long time stars with a large speed (large proper motion and/or large radial velocity) have been selected to find metal-poor stars \citep{roman1950}.
In fact, stars in the solar vicinity with a high velocity (with respect to the Sun) surely belong to the Galactic halo and as a consequence they are metal-poor.
With the Gaia releases, the kinematic selection of halo stars to catch metal-poor stars has regained popularity \citep[see e.g.][]{ghs1,mataspinto22,ghs2_2024}.
A detailed kinematical knowledge has also been widely used to connect a star to a particular stream \citep[see e.g.][]{ibata2024}.
These streams, witnesses of accreted dwarf galaxies or disrupted clusters \citep{martin2022}, are necessary to reconstruct the past history and the formation 
of the Milky Way.

The total velocity with respect to the Galactic centre of the star depends on its position in the Galaxy (coordinates and parallax), the proper motions and
the radial velocity ($\rm V_r$), that is derived from the RVS spectra \citep{Sartoretti}.
Each quantity is provided with an uncertainty and other flags (e.g. the renormalised unit weight
error,  ruwe) are a useful guide to decide what  Gaia data are trustable.

In the Gaia\,DR3, we selected a sample of stars with $\rm |V_r|>500$\,\kms, to be observed at ESO to confirm/refute the high radial velocity.
A sample of 45 stars have been observed and the encouraging results (all are indeed high $\rm V_r$ stars) is presented in \citet{rvs2}.
We here present the following sample of 26 high $\rm V_r$ stars observed with UVES at VLT.

\section{Observations} 

The UVES spectra were observed in programme 110.23V0.001 and programme 112.25D7.001, with 
setting DIC2\,437+760 (wavelength ranges 373--499 and 565--946\,nm),
slit 0\farcs{4}\ in the blue arm and 0\farcs{3}\ in the red arm (resolving power of 80\,000\, in the blue arm and 110\,000\ in the red arm).
With this high resolution the doppler width of the lines is completely resolved and the width of the lines
is dominated by macroturbulence and stellar rotation.
We reduced all the spectra with the ESO pipeline (on gasgano version \footnote{\url{https://www.eso.org/sci/software/legacy/gasgano.html}}).

\section{Analysis}

\begin{figure}[t]
\centering
 \includegraphics[width=\hsize,clip=true]{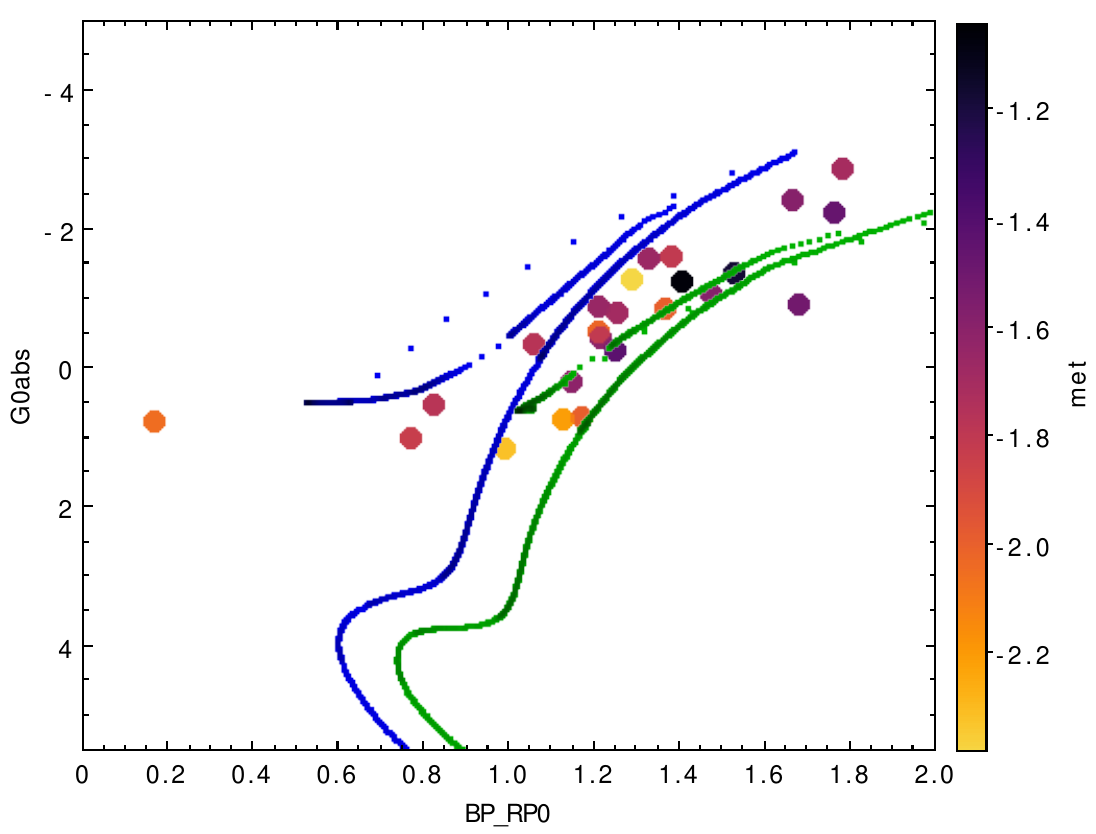}
\caption{The colour-magnitude diagram of the 25 stars with ``correct'' radial 
velocity determination from Gaia\,DR3 compared to a BASTI \citep{pietrinferni2021}
$\alpha$-enhanced isochrones of 12\,Ga of metallicity $-0.60$ (green) and  $-1.55$ (blue)}
\label{fig:iso}
\end{figure}

The sample consists of 26 stars selected to confirm/dismiss the high radial velocity
($\rm |V_r|>500$\,\kms) derived by Gaia\,DR3.
All the stars are confirmed high radial velocity, except RVS929, which shows a complex spectrum.
The high radial velocity stars are all evolved (red giant branch or asymptotic giant branch stars) and metal-poor,
except RVS1383, that is a metal-poor horizontal branch or sub-giant star.
In Fig.\,\ref{fig:iso} 25 of the stars (all except RVS929) are compared to BASTI \citep{pietrinferni2021} $\alpha$-enhanced 
isochrones of   12\,Ga and metallicities $-0.60$ and  $-1.55$.
It may be noticed that  isochrones imply higher metallicities
than our spectroscopically determined values. One could suspect that there is
an offset between spectroscopic [Fe/H] and the metallicity used in the
isochrone computation. However, there is no reason why such an offset should arise,
note also that the BASTI isochrones use the same set of solar abundances as we do.
We believe instead that the reason is more subtle and it is linked to the 
treatment of convection in the theoretical evolutionary models.
Like most isochrones, BASTI treats convection using the mixing length theory,
since there is no way to determine the mixing length from first
principles  \citet{pietrinferni2021} adopt $\alpha_{ML} = 2.006$,
from a solar model calibration. 
The stellar radii and effective temperatures, but not the luminosities, 
depend on the adopted mixing length. It is well known that a lower mixing length 
moves the red-giant branch to lower effective temperatures 
\citep[see][figure 1]{castellani1999}.
Our interpretation of these observations is 
that these metal-poor giants would be better represented by isochrones
computed with a lower mixing length. It is also not obvious that the same
mixing length is suitable for all stars. 
In fact more physically motivated models rely on radiation hydrodynamical simulations
that provide entropy as a function of 
effective temperature, surface gravity, and chemical composition, 
these provide a mean to calibrate the adiabatic convection
\citep[see][and references therein]{manchon2024}. 

\subsection{Radial velocity} \label{sec:vr}

\begin{figure}[t]
\centering
 \includegraphics[width=\hsize,clip=true]{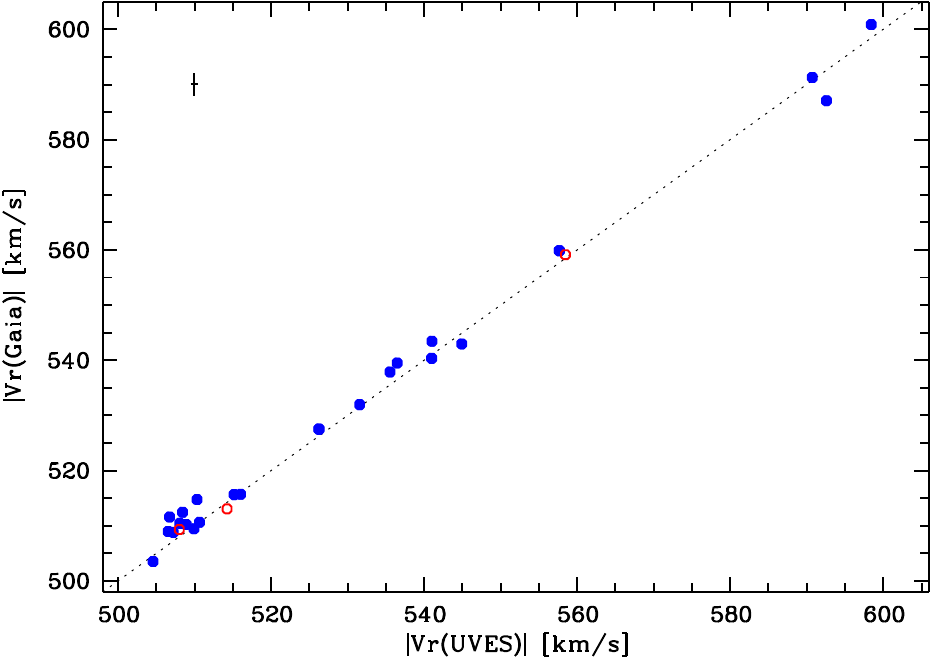}
\caption{Comparison of the modulus of the
radial velocity derived by the UVES spectra with the one from Gaia\,DR3 for the stars with confirmed high radial velocity.
For ease of representation, 
the absolute radial velocities are reported. The blue filled circles are positives radial velocities and
open red circles the negative ones. A representative uncertainty is reported on the top left corner.}
\label{fig:vrad}
\end{figure}

We derived the radial velocity with our own template matching code, 
using the wavelength region 840\,nm -- 880\,nm, that is roughly equal to the RVS wavelength range.
We made this choice to avoid any systematic difference
between the UVES radial velocities and the Gaia radial velocities that could
be due to different lines being used. 
The templates were computed with the SYNTHE code \citep{K05}
using ATLAS 9 model atmospheres \citep{K05} using the Opacity Distribution
Functions of Mucciarelli et al. (in prep.) with microturbulence 1\,\kms\
and $\alpha$ element enhancement +0.4\,dex.
More details on the radial velocity investigation and on the kinematics of the stars are provided in Katz et al. (to be sub.).
In Table\,\ref{coord} we report the Gaia\,DR3 radial velocities for all the stars except RVS929, while 
in Fig.\,\ref{fig:vrad} the modulii of the radial velocities are compared.
The uncertainty in the radial velocity determination on the UVES spectra is dominated by the position of the star in the slit, 
and this is of the order of 0.5\,\kms.

\subsection{Stellar parameters and chemical analysis} \label{sec:param}

We derived the stellar parameters as described by \citet{lombardo21}, using the Gaia\,DR3 \citep{gaiadr3} photometry and parallax:
comparing the $G_{BP}-G_{RP}$ Gaia colour to a grid of synthetic colours in order to derive the effective temperature, \teff.
Once \teff\ was obtained, we derived the gravity from the Gaia parallax, corrected by the zero-point as suggested by \citet{lindegren21}, by using the Stefan-Boltzmann equation.
We used $-1.5$ as first guess metallicity and a mass of $\rm 0.8\,M_\odot$. 

Extinction values were estimated by integrating from the Sun to the target star within the three-dimensional (3D) extinction density maps from \cite{Vergely22}. The maps are based on the tomographic inversion of a large amount of extinction-distance pairs of individual targets, following a hierarchical technique. The latest maps used extinctions of $\simeq$ 35 500 000 individual stars, derived from Gaia G, GBP, and GRP and 2MASS J, H, and KS photometric data, and their distances from Gaia EDR3 parallaxes. Accurate extinctions of $\simeq$ 6 000 000 stars derived from both spectroscopic and photometric measurements were added and the two catalogues were inter-calibrated following a newly devised technique \citep[see][]{Vergely22}. The achievable spatial resolution of the maps is governed by the target star density and distribution, which is highly variable and depends primarily on the distance to the Sun. The hierarchical inversion technique ensures that the spatial correlation kernel is compatible everywhere with the local target density. For reasons linked with the computational time, several maps with different maximum spatial resolutions and distance coverages were built (see the G-TOMO online facility at \url{https://datalabs.esa.int)}. Here, we used a composite integration starting within the highest resolution map, and, if the line-of-sight reaches the boundary of the map, continuing the integration in the lower resolution (and wider) map, and so on. The maps are limited to a distance to the Plane Z of 400 pc. For targets with Z$\geq$400 pc, the extinction beyond Z=400 pc is considered negligible.

With the first-guess parameters, we ran \mygi\ \citep{mygi14} to derive the metallicity that was used to update the parameters.
With the final parameters we computed an ATLAS\,9 models \citep{K05} for each star using the ODFs of Mucciarelli et al. (in prep.).
With this model, we computed a grid of synthetic spectra with Turbospectrum \citep{alvarez1998,turbo}, in steps of 0.2\,dex in the abundance of all elements around the metallicity of the model.
The grid was used by \mygi\ to derive the detailed chemical abundances, the same procedure has been used in \citet{caffau_leo_2024}.
Several stars in the sample have a signal-to-noise ratio (S/N) too low to allow us to detect weak \ion{Fe}{i} lines in order to derive the microturbulence
from the flat trend of A(Fe) versus equivalent width.
To be homogeneous in the analysis and having a sample of metal-poor giant stars, we adopted the microturbulence provided by the calibration of \citet{mashonkina17},
except for RVS1383, an early F-type  star, for which we adopted 2\,\kms.

The stars here investigated have stellar parameters similar to the stellar samples analysed by \citet{rvs2} and \citet{mataspinto22} as the chemical investigation
was performed in a similar way:
for systematic uncertainties and NLTE investigations refer to these papers.

The majority on the stars in the sample are not known in the SIMBAD astronomical database\footnote{\url{}https://simbad.cds.unistra.fr/simbad/} and just one star (RVS1315) has a complete chemical analysis in the literature.

\subsection{Remarks on individual stars}\label{secabbo}

Details on the stars can be found in Table\,\ref{tab:stars},
the identifications and coordinates are provided in Table\,\ref{coord} and  
full details about the individual chemical abundances derived for each star and the
spectral lines used are made available at CDS.
Here below some remarks on few stars.

\subsubsection{RVS918}
\begin{figure}
\centering
\includegraphics[width=\hsize,clip=true]{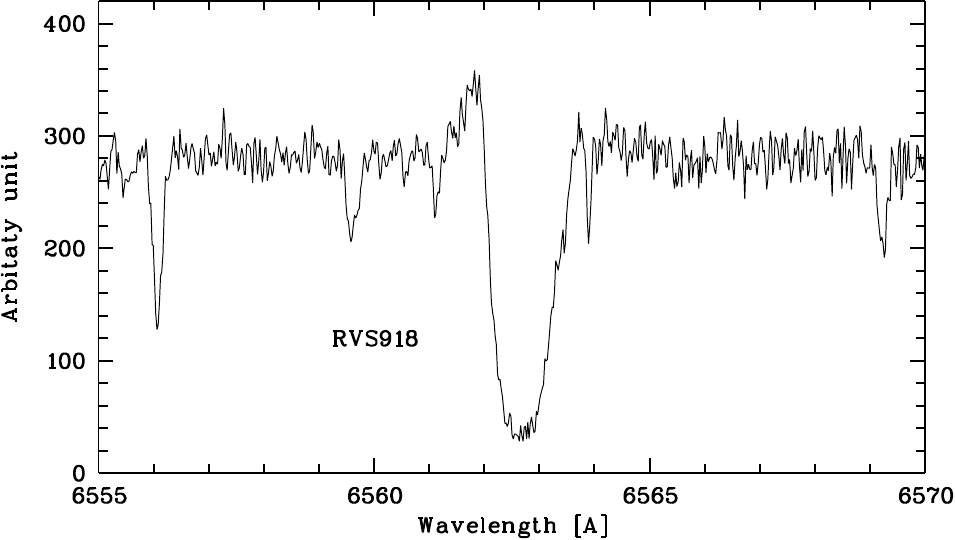}
\caption{The observed spectrum of RVS918 in the wavelength range of $\rm H\alpha$.}
\label{fig:obsrvs918}
\end{figure}

RVS918 (Gaia DR3 5218683236590712192,   2MASS J09142571--7235089)
is a Long-Period Variable Candidate in the catalogue by \citet{lebzelter2023}.
H$\alpha$ shows an inverse P-cygni profile (see Fig.\,\ref{fig:obsrvs918}).

\subsubsection{RVS920}
RVS920 has a \ion{Li}{i} at 607\,nm providing in LTE $\rm A(Li)\sim 0.8$.
This star is too cool (\teff=4732\,K and \logg=2.35, see Table\,\ref{tab:stars}) to be on the Mucciarelli plateau \citep{mucciarelli21} 
and in fact A(Li) is below the value of this plateau.
The star is probably in the phase of diluting Li in its photosphere.

\subsubsection{RVS924}
RVS924 (Gaia DR3 5642811738112717824) is known in SIMBAD and it is in a crowded area with other 5 objects within 20 arcsec.
It is a carbon enhanced metal-poor \citep[CEMP, see ][and references therein, for a definition and discussion on this class]{BCFS} star, with $\rm [Fe/H]\sim -2$ and $\rm [C/Fe]\sim 1.5$.
The S/N is too low (13 and 24 at 450 and 650\,nm) to be able to derive, with a good accuracy, detailed abundances.

\subsubsection{RVS928}
\begin{figure}
\centering
\includegraphics[width=\hsize,clip=true]{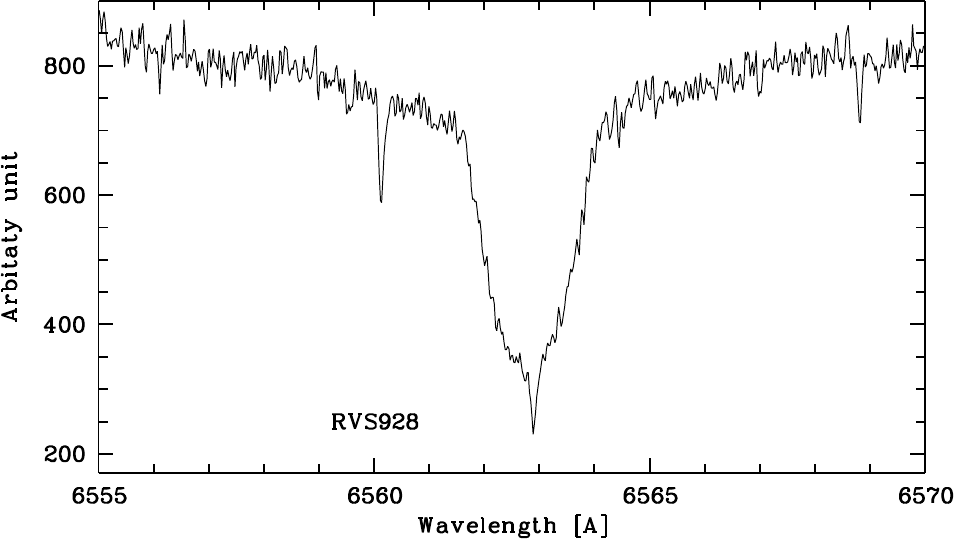}\\
\includegraphics[width=\hsize,clip=true]{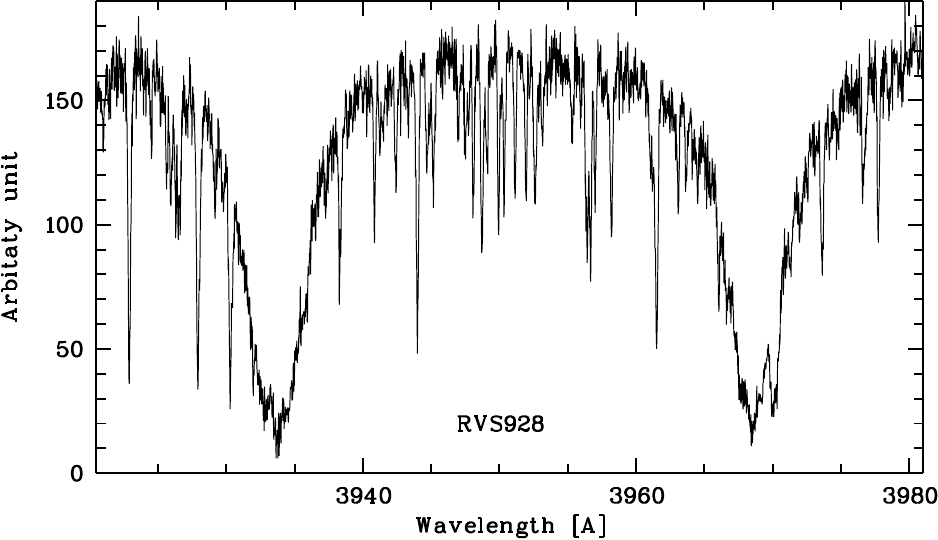}
\caption{The observed spectrum of RVS928, upper panel: in the wavelength range of $\rm H\alpha$, lower panel: in the wavelength range of \ion{Ca}{ii}-H and -K lines.}
\label{fig:obsrvs928}
\end{figure}

RVS928 (Gaia DR3 3778236623817788416) has broad lines, we estimated
a width of about 10 \kms , this could be interpreted
as rotational broadening. It has a  narrow absorption line in the H$\alpha$ core (see Fig.\,\ref{fig:obsrvs928}, upper panel),
that may suggest the presence of a cooler companion, however in the rest of the spectrum
we cannot detect any sign of the spectrum of a secondary star. 
This star appears too cool to be classic F-type shell star, such as  $\pi^2$\,Peg \citep{slettebak1986}, however this narrow absorption core
is suggestive of a cool circumstellar shell. 
The shell is confirmed by the presence of a red-shifted component of the \ion{Ca}{ii}-H line (see Fig.\,\ref{fig:obsrvs928}, lower panel) and by the very strong D \ion{Na}{i} lines.

\subsubsection{RVS929}
\begin{figure}
\centering
\includegraphics[width=\hsize,clip=true]{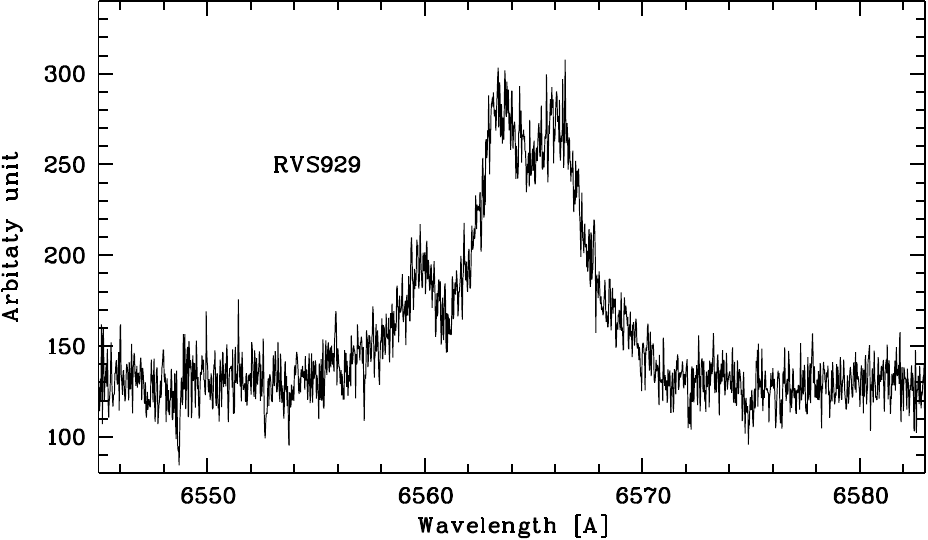}
\caption{The observed spectrum of RVS929 in the wavelength range of $\rm H\alpha$.}
\label{fig:obsrvs929}
\end{figure}

\begin{figure}
\centering
\resizebox{7.5cm}{!}{\includegraphics{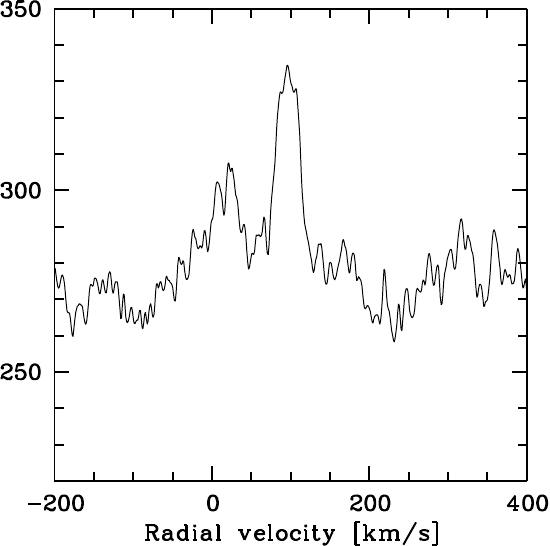}}
\caption{The cross-correlation function of the UVES spectrum of RVS929 in the range 
570\,nm - 650\,nm with a synthetic spectrum of solar metallicity, effective temperature of 5400\,K
and log g = 4.5.}
\label{fig:xc_rvs929}
\end{figure}

RVS929 (Gaia DR3 5313687432150359424) is unknown in SIMBAD. 
The $\rm H\alpha$ shows three wide and strong emissions
(see Fig.\,\ref{fig:obsrvs929}) at radial velocity (blue to red)
--135 \kms , +17 \kms\ and +134 \kms.
A cross correlation with a template close to the parameters
used by the Gaia DR3 to determine the radial velocity, over the 
wavelength range 570\,nm to 650\,nm, thus excluding H$\alpha$,
shows at least two wide peaks, shown in Fig.\,\ref{fig:xc_rvs929},
that corresponds to radial velocities of 15 \kms\ and 96 \kms.
The \ion{Ca}{ii} IR triplet lines show an emission at radial
velocity of --110 \kms and an absorption at 97 \kms.
This suggests that the system is an SB2, although a higher
multiplicity is possible. The two reddest peaks of the H$\alpha$
may be associated to the systems of absorption lines, while
the bluest H$\alpha$ emission seems to share the 
same radial velocity as the \ion{Ca}{ii} IR triplet emission lines. 
Given that the Gaia RVS range is plagued by these emission
lines it is unsurprising that the Gaia radial velocity is off.
The system is quite complex with at least two stellar components
and possibly a circumstellar or disc component.

\subsubsection{RVS933}
RVS933 (Gaia DR3 5358307297639362304) is defined as a high velocity star in SIMBAD and the star is listed in the hypervelocity sample by \citet{parthasarathy2024}.

\subsubsection{RVS1280}
RVS1280 (Gaia DR3 2966726811218049280), known also as BD--18 1229, is classified as a long-period variable in SIMBAD.

\subsubsection{RVS1281}
RVS1281 (Gaia DR3 5325632011075073408) is a high velocity star, probably belonging to the moving group LRK2022 \citep{LRK2022},
that includes other  74 stars.
The star is  in the catalogue of hypervelocity stars presented by \citet{parthasarathy2024}. 
The star is cooler than our grids and the S/N is low.
Assuming a temperature of 3500\,K, we estimate a metallicity of about $-2$.

\subsubsection{RVS1315}
With parameters \teff = 4327 and \logg=1.20, for RVS1315 (Gaia DR3 5717948445741886720, BD--16 2232)
we derived $\rm [Fe/H]=-1.80\pm 0.13$ (see Table\,\ref{tab:stars}), 
$\rm [Y/Fe]=-0.03$ and $\rm [Eu/Fe]=0.58$.
The star has been investigated by \citet{aguado2021} and, adopting the 
parameters \teff = 4503 and \logg\ of 1.27, they derived: 
$\rm [Fe/H]=-1.57\pm 0.12$, $\rm [C/Fe]=-0.46\pm 0.10$, $\rm [Sr/Fe]=-0.44\pm 0.10$, $\rm [Y/Fe]=-0.14\pm 0.10$,
$\rm [Ba/Fe]=-0.14\pm 0.09$, $\rm [Eu/Fe]=0.57\pm 0.08$.
From the kinematics they classified this star as a member of the Gaia-Sausage-Enceladus accretion event
\citep{belokurov18,haywood18,helmi18}.

\subsubsection{RVS1383}
RVS1383 (Gaia DR3 4821430166509061632) also known as TYC 7067--549--1,  was classified as a BHB star by \citet{culpan2021}.
We derive a projected rotational velocity ($\rm V\sin{i}$) of about 12\,\kms\ from the optical lines in the UVES spectrum.
The position in the colour-magnitude diagram, compared to BASTI isochrones at metallicity --1.9, is compatible with the star being: 
(i) an evolved blue straggler or a genuine young star (1--2\,Ga and $\rm 1.5 M_\odot$), in this case the star is in its turn-off or (ii) an old ($\sim 12$\,Ga and $\rm 0.8 M_\odot$) BHB star.
The derived $\rm V\sin{i}$ is not a discriminating factor: this value is compatible with both cases \citep{billi2024}. 

Due to the extreme kinematics, it can in fact be a real metal-poor star, perhaps recently formed in the interaction of a
dwarf galaxy with the Milky Way \citep{preston2000,hammer2024} and then accreted to the Galaxy.
No Li is detectable, but the Li feature would be anyway too weak to be detectable.
The lines used to analyse this star are different from the lines investigated in the other, cool stars.
The line-to-line scatter in the abundances are small, except for \ion{Ti}{ii}:
the 36 lines  fall in two groups: (1) 16 lines with $\rm\langle A(Ti)\rangle =3.46\pm 0.06$ and (i) 14 lines with $\rm\langle A(Ti)\rangle =2.78\pm 0.07$.
This is not related to the choice on the broadening or the micro-turbulence.
We keep the lines of the ``high'' A(Ti). 

\subsubsection{RVS1392}
RVS1392 (Gaia DR3 2921543205513614208) is a high velocity star
belonging to the LRK2022 moving group \citep{LRK2022}.
The star is highlighted as a high radial-velocity star by \citet{parthasarathy2024}.
This star has a visible Li feature, that provide in LTE A(Li)=1.13, compatible with the Mucciarelli plateau \citep{mucciarelli21}.

\begin{table*}
\caption{Stars.}
\label{tab:stars}
\begin{tabular}{lrrrlrrlll}
\hline
\smallskip
Star & \teff\  & \logg\ & $\xi$ & [Fe/H] & S/N & S/N & Li & \ion{Ca}{ii}-H and -K & Note \\ 
     &         &        &       &        & 450\,nm & 650\,nm & & \\
\hline
RVS911  & 3830 & 0.61 & 1.85 & $-1.47 \pm 0.15  $ &  29 &  78 & no  & & \\
RVS914  & 4500 & 1.43 & 1.87 & $-2.38 \pm 0.12  $ &  25 &  54 & no  & & \\
RVS916  & 4597 & 1.64 & 1.75 & $-1.65 \pm 0.14  $ &  31 &  55 & no  & tiny emission & \\
RVS917  & 4704 & 2.12 & 1.57 & $-1.60 \pm 0.15  $ &  24 &  43 & no  & & \\
RVS918  & 3954 & 0.62 & 1.90 & $-1.58 \pm 0.15  $ &   7 &  26 & no  & & H$\alpha$ inverse P-cygni profile \\
RVS919  & 4500 & 1.85 & 1.62 & $-1.42 \pm 0.14  $ &  19 &  35 & no  & & \\
RVS920  & 4732 & 2.35 & 1.49 & $-2.21 \pm 0.13  $ &  24 &  41 & yes & & \\
RVS922  & 4905 & 2.00 & 1.75 & $-1.76 \pm 0.13  $ &  18 &  29 & no  & & \\
RVS923  & 4079 & 1.15 & 1.69 & $-1.17 \pm 0.17  $ &   8 &  26 & no  & & \\
RVS924  & 4641 & 2.30 & 1.45 & $-2.0  \pm 0.30  $ &  13 &  24 & no  & & CEMP-r/s \\
RVS926  & 4162 & 1.32 & 1.72 & $-1.58 \pm 0.16  $ &   3 &  26 & no  & & poor S/N \\
RVS928  & 5674 & 2.83 & 1.75 & $-1.83 \pm 0.12  $ &  45 &  51 & no  & & narrow absorption on H$\alpha$ \\
RVS929  &      &      &      &                    &     &     &     & &  H-$\alpha$ complex structure \\
RVS930  & 5512 & 2.58 & 1.77 & $-1.77 \pm 0.17  $ &  11 &  15 & no  & & \\
RVS931  & 4271 & 1.31 & 1.70 & $-1.04 \pm 0.16  $ &  15 &  30 & no  & & \\
RVS932  & 3875 & 1.17 & 1.60 & $-1.51 \pm 0.17  $ &   6 &  18 & no  & & poor S/N \\
RVS933  & 4337 & 1.50 & 1.72 & $-2.00 \pm 0.14  $ &  15 &  44 & no  & & \\
RVS937  & 4510 & 1.64 & 1.72 & $-1.69 \pm 0.13  $ &  26 &  48 & no  & tiny emission & \\
RVS938  & 4591 & 1.78 & 1.71 & $-2.02 \pm 0.15  $ &  23 &  39 & no  & & \\
RVS941  & 4411 & 1.26 & 1.84 & $-1.66 \pm 0.16  $ &  29 &  53 & no  & & \\
RVS1280 & 3866 & 0.37 & 1.99 & $-1.70 \pm 0.16  $ &  51 &  96 & no  & emission & \\
RVS1281 & 3500 & 0.50 &      & $\sim -2$        &   5 &  30 &     &  & cool star \\
RVS1315 & 4327 & 1.20 & 1.83 & $-1.80 \pm 0.13  $ &  85 &  98 & no  & tiny emission & analysed by \citet{aguado2021} \\
RVS1358 & 4575 & 1.81 & 1.65 & $-1.60 \pm 0.14  $ &  59 &  85 & no  & tiny emission \\
RVS1383 & 8100 & 3.60 & 2.00 & $-2.05 \pm 0.08  $ &  99 & 100 & no  & & hot star \\
RVS1392 & 5038 & 2.65 & 1.54 & $-2.31 \pm 0.10  $ &  77 &  93 & yes & & \\
\hline
\end{tabular}
\end{table*}

\begin{figure}
    \centering
    \resizebox{7.4cm}{!}{\includegraphics{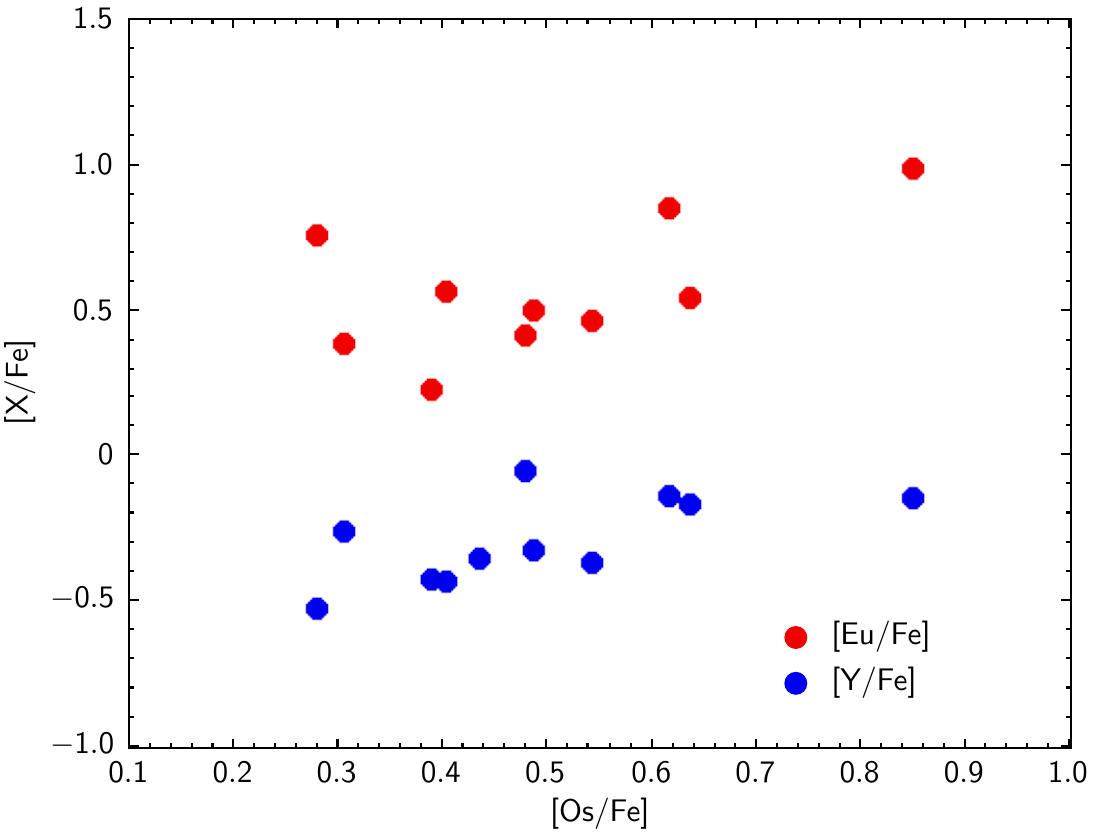}}
    \caption{[Y/Fe] (blue dots) and [Eu/Fe] (red dots) as a function of [Os/Fe].}
    \label{fig:Os_Y_Eu}
\end{figure}

\begin{figure}
    \centering
    \resizebox{7.4cm}{!}{\includegraphics{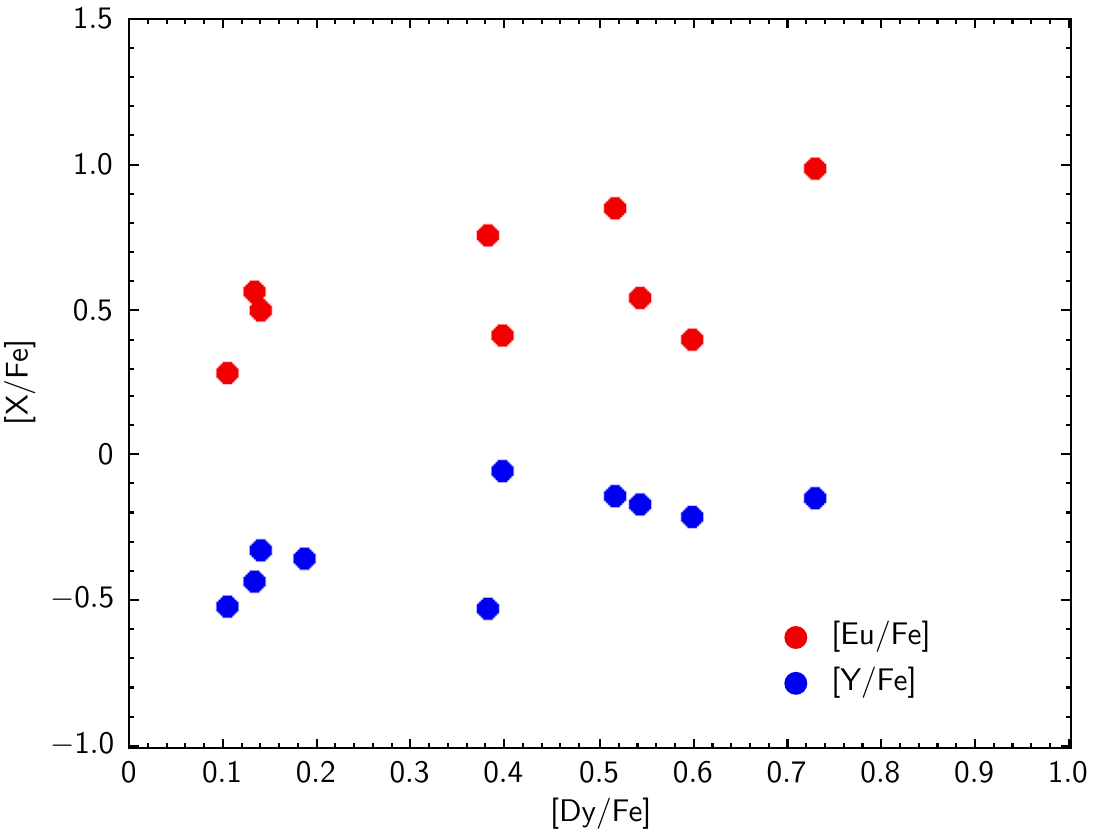}}
    \caption{[Y/Fe] (blue dots) and [Eu/Fe] (red dots) as a function of [Dy/Fe].}
    \label{fig:Dy_Y_Eu}
\end{figure}

\begin{figure}
    \centering
    \resizebox{7.4cm}{!}{\includegraphics{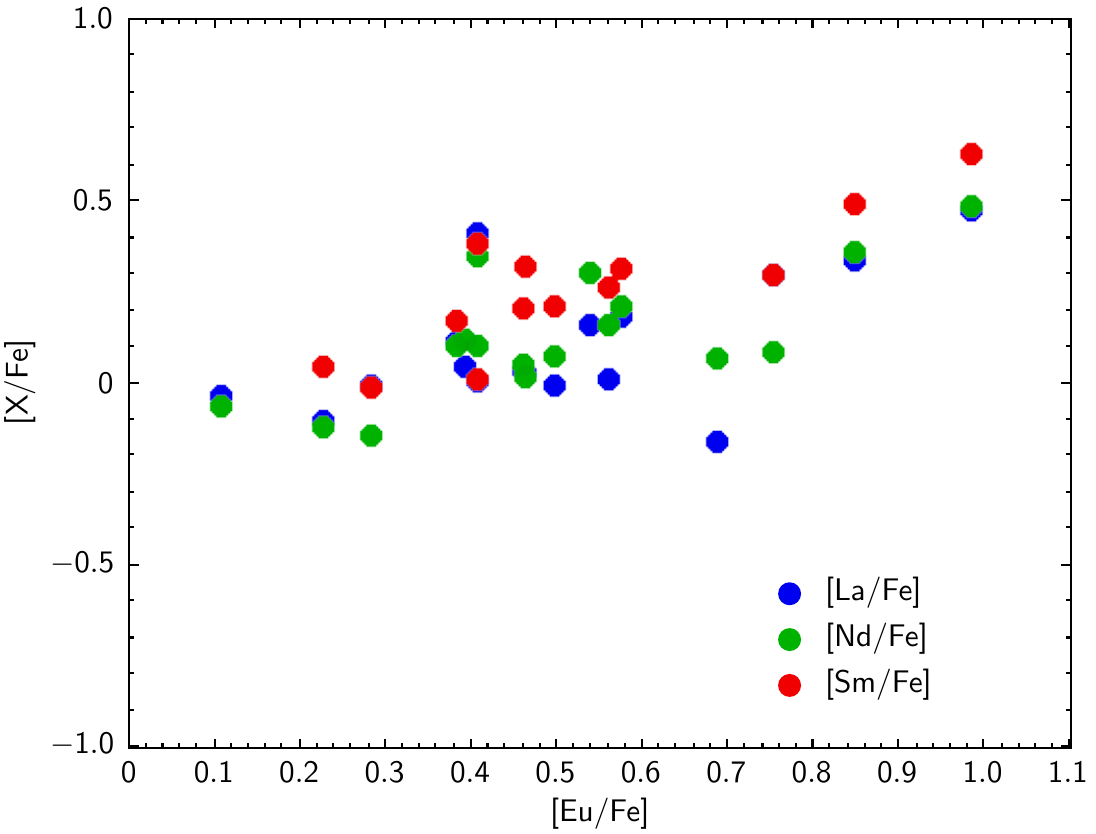}}
    \caption{[La/Fe] (blue dots), [Nd/Fe] (green dots) and [Sm/Fe] 
    (red dots) as a function of [Eu/Fe].}
    \label{fig:Eu_La_Nd_Sm}
\end{figure}

\section{Neutron capture elements}

The wide wavelength coverage and good S/N ratios of many
of our spectra allowed us to determine a rich inventory of abundances of
neutron capture elements ranging from Rb to Os.
Among these some seldom analysed elements in the metal-poor regime: 
four measurements for Rb, 19 measurements for Ce, 17 for Sm, 12 for Dy and 11 for Os.
All elements with $Z > 30$ are formed through some neutron capture process, 
the two main ones are the slow neutron capture process, hereafter s-process, 
in which each neutron capture is followed by a $\beta-$decay 
\citep[see][and references therein]{arcones_origin_2023} and
the rapid neutron capture process, hereafter r-process, in which
several neutrons are captured before $\beta-$decay occurs
\citep[see][ and references therein]{cowan_origin_2021}.
All of our stars are enhanced in the pure r-process element Eu, with seven out of 17
stars with measured Eu, with [Eu/Fe]$> 0.5$. This situation is common
in this metallicity regime \citep{MINCEII,MINCEIII}. 
Osmium is considered to be mainly formed through the rapid neutron capture process, and
in Fig.\,\ref{fig:Os_Y_Eu} we show  [Y/Fe] and [Eu/Fe] as a function of [Os/Fe].
By using the ratios of the elements to iron we attempt to remove the metallicity effects.
Yttrium is a pure s-process product, while Eu is a pure r-process product. 
Neither of the correlations is statistically significant\footnote{We consider a correlation to be
statistically significant if its probability, estimated from Kendall's $\tau$ is larger than 95\%.}, 
however
while the correlation with [Y/Fe] has a probability of 94.8\%,
as estimated from Kendall's $\tau$, that with [Eu/Fe] has a probability of only 87\% .
This suggests that 
in these stars the Os present in the gas at the time of their
formation has been formed both by the s-process and by the r-process, since there
is no clear correlation with the products of either.
Among the two the s-process seems to be favoured. 
The situation is reversed with Dy, another element that is expected
to be formed mainly by the r-process. 
In Fig.\,\ref{fig:Dy_Y_Eu} we show [Y/Fe] and [Eu/Fe] as a function
of [Dy/Fe]. In this case the probability of correlation with [Y/Fe] is at 86.8\% ,while
that with [Eu/Fe] is at 94.8\% .
Again we interpret this as evidence that both s- and r- processes are at work, 
in this case the r-process seems to be favoured.

Exploring all the correlations of [X/Fe] with [Y/Fe],
we find significant correlations with [La/Fe], 99.5\%, 
[Nd/Fe], 99.8\%.
With [Eu/Fe] we find statistically
significant correlations for [La/Fe], 97.9\%, [Nd/Fe], 98.7\%,  and
[Sm/Fe], 99.8\% .

In Fig.\,\ref{fig:Eu_La_Nd_Sm} we show the abundance ratios of the
three elements with statistically correlations with Eu.
Both the correlation and the scale seem to be the same, while
La and Nd also show a statistically significant correlation with Y, Sm does not
(probability 54\%).
This may suggest that Sm has been produced only by the r-process.

\section{Discussion and conclusions}

\begin{figure}
\centering
\includegraphics[width=\hsize,clip=true]{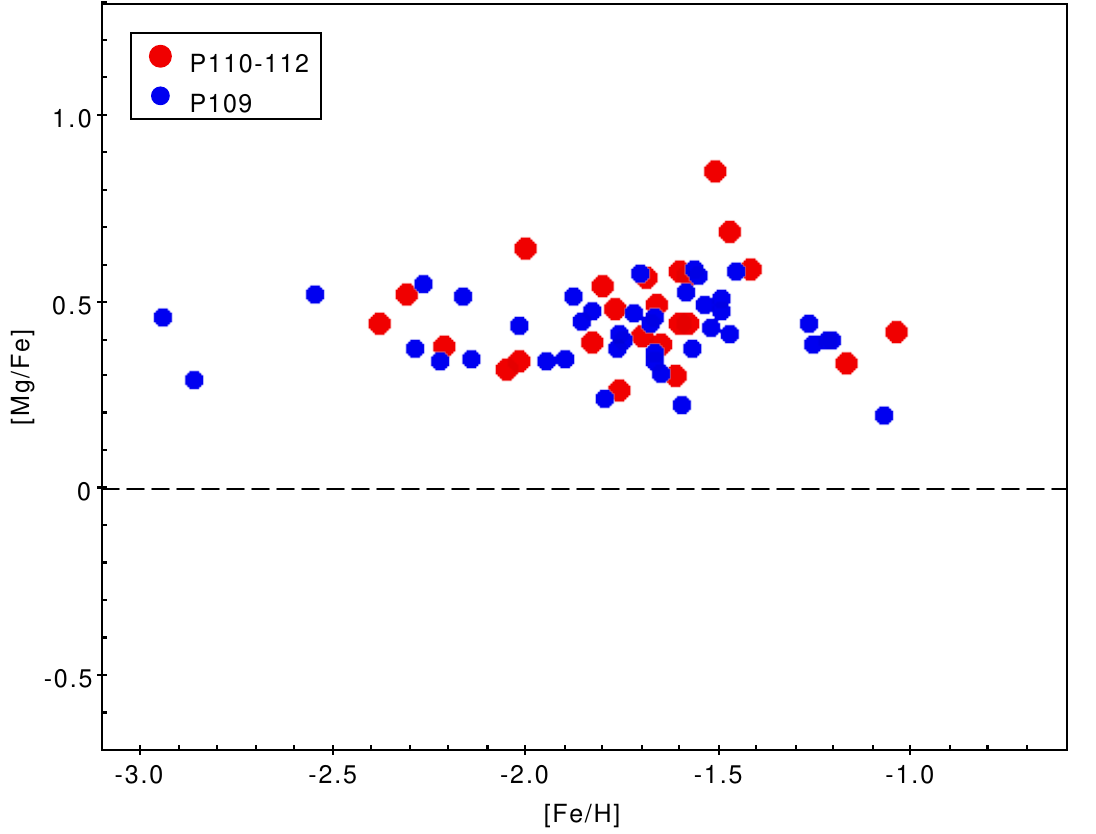}
\caption{[Mg/Fe] vs. [Fe/H] for the sample of stars (red filled circles) compared to the sample observed in P109 \citep[][blue filled circles]{rvs2}.}
\label{fig:mgfe}
\end{figure}

All the stars in the sample, except RVS929, have high absolute radial velocity indeed, as measured by Gaia.  We then confirm the results provided in Gaia\,DR3.
The stars with confirmed high absolute radial velocity  
are metal-poor and span the metallicity range --1 to --2.4.
They are all $\alpha$-enhanced as expected at this metallicity regime.
We derived $\rm\langle[Mg/Fe]\rangle =0.48\pm 0.13$, $\rm\langle[Si/Fe]\rangle =0.40\pm 0.15$,
$\rm\langle[Ca/Fe]\rangle =0.26\pm 0.10$
(see e.g. in Fig.\,\ref{fig:mgfe} [Mg/Fe] versus [Fe/H] in comparison with the previous sample of high $\rm V_r$ stars analysed in \citealt{rvs2}).
Sulphur abundance was derived only in nine stars and is based on the \ion{S}{i} lines of Mult.\,1 (920\,nm) which are affected by
strong NLTE corrections. The wavelength range is also badly contaminated by telluric absorption, so the high [S/Fe] ratio and its large star-to-star
scatter are not as trustable as for the other $\alpha$ elements.
Oxygen, based on the [OI] lines, is also enhanced ($\rm\langle[O/Fe]\rangle =0.53\pm 0.28$, using A(Fe) from \ion{Fe}{ii} lines). 
For the hot star RVS1383, that stands out with an extremely high [O/Fe] of 1.71, A(O) is based on the 777\,nm \ion{O}{i} lines and the line at 844\,nm, all strongly sensitive to NLTE effects. This star shows a sub-solar $\rm [Si/Fe]=-0.13$.

Two stars have a high [Na/Fe] ratio (RVS928 and RVS930), for both based on the D double lines. Neither of the two has a measurable O abundance, 
it is thus likely that their O abundance is not high, lower than  that of the other stars of the sample with similar parameters. 
A high Na and low   O is a typical signature of second generation stars in Globular Clusters \citep{gratton01,bl18}.
This fact may suggest that these two stars have escaped from a Globular Cluster. 
The two stars show a Mg abundance that is in line with the other stars.
Of the two only RVS928 has a measured Al abundance and it is in line
with that of the others. Thus if these stars are second generation stars
escaped from a globular cluster, it must have been a low mass one,
not showing the Mg-Al anticorrelation \citep{pancino17}. 
 
\begin{figure}
\centering
\includegraphics[width=\hsize,clip=true]{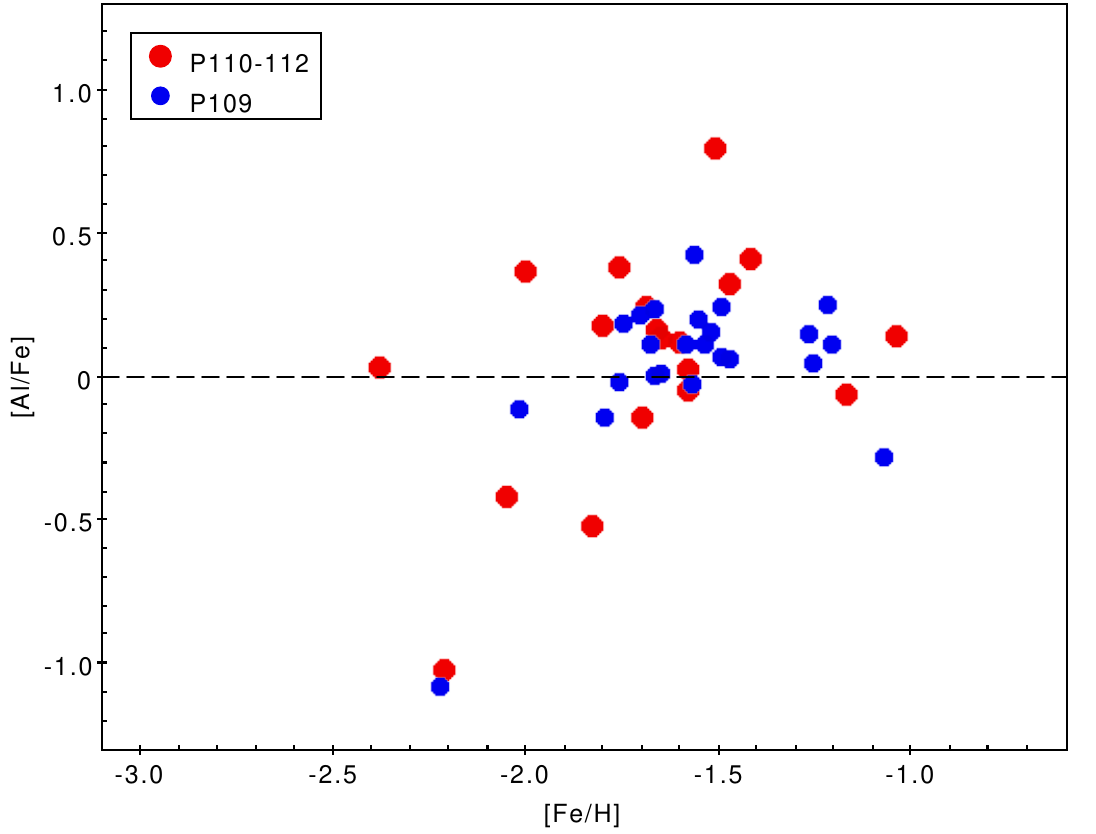}
\caption{[Al/Fe] vs. [Fe/H] for the sample of stars (red filled circles) compared to the sample observed in P109 (blue filled circles).}
\label{fig:alfe}
\end{figure}

The [Al/Fe] ratio, as in the previous sample \citep{rvs2}, spans from $-1.0$ to $+0.8$ (see Fig.\,\ref{fig:alfe}).
A(K) is derived in 18 stars by investigating the strong 768.8\,nm \ion{K}{i} line.
A(Sc) from neutral and ionsed lines are in reasonable agreement ($\rm\langle[Sc/Fe]\rangle$ of $-0.14\pm 0.07$ and $-0.01\pm 0.13$, respectively)
and the same is true for Ti ($\rm\langle[Ti/Fe]\rangle$ of $0.23\pm 0.08$ and $0.22\pm 0.10$, respectively).
Iron peak elements are close to iron, except Mn with an underabundance ($\rm\langle[Mn/Fe]\rangle=-0.32\pm 0.10$) surely also related to NLTE effects
\citep[see e.g.][]{bergemann19mn} and Cu ($\rm\langle[Mn/Fe]\rangle=-0.45\pm 0.16$) as well also related to NLTE effects.

\begin{figure}
\centering
\includegraphics[width=\hsize,clip=true]{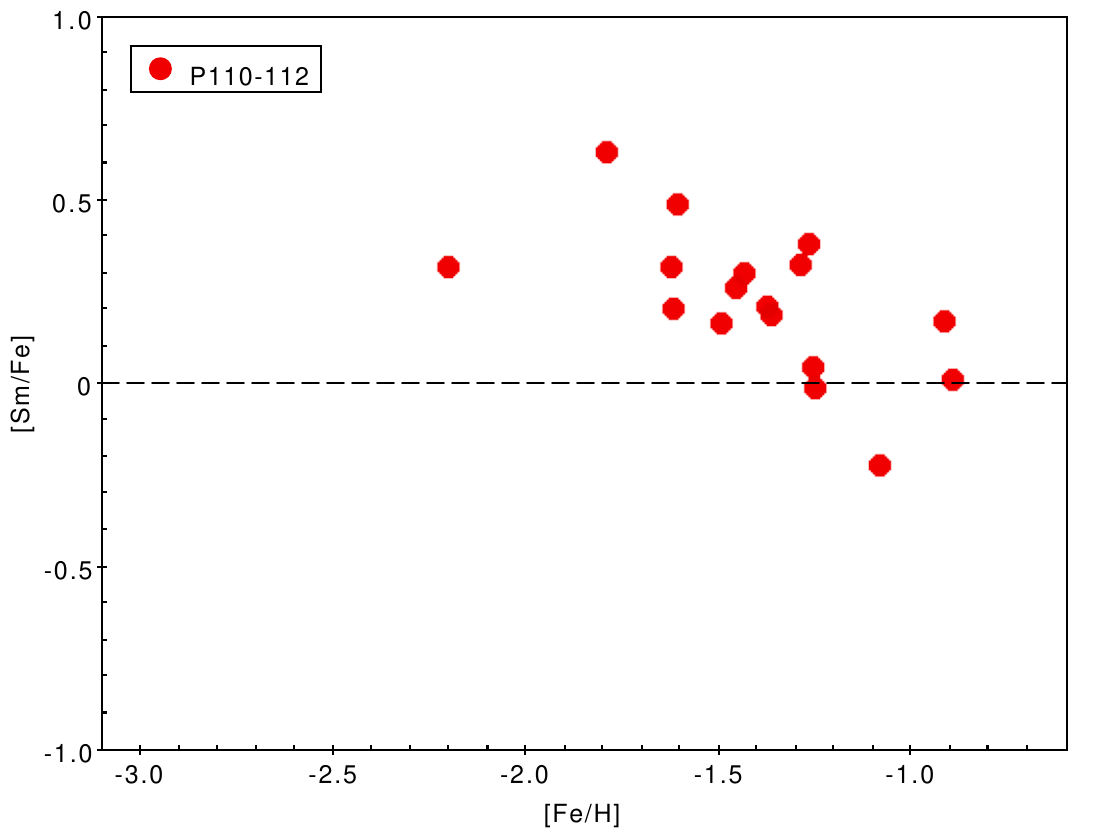}
\caption{[Sm/Fe] vs. [Fe/H] for the sample of stars (red filled circles).}
\label{fig:smfe}
\end{figure}

\begin{figure}
\centering
\includegraphics[width=\hsize,clip=true]{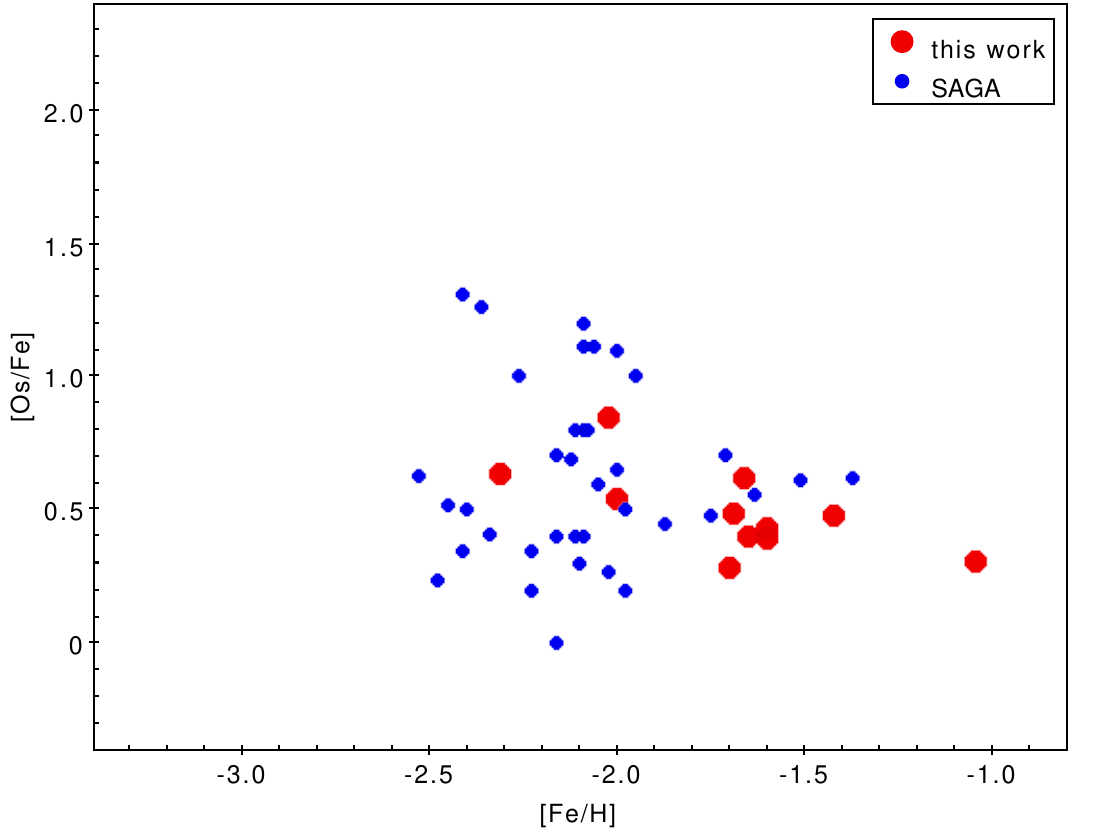}
\caption{[Os/Fe] vs. [Fe/H] for the sample of stars (red filled circles) compared to the stars in the SAGA database \citep[][blue filled circles]{saga2008}.}
\label{fig:osfe}
\end{figure}

The heavy elements show a larger star-to-star scatter than the lighter elements as found
by other investigations in this metallicity regime \citep[see e.g.][]{MINCEII,MINCEIII}.
We derived Sm for 17 stars (see Fig.\,\ref{fig:smfe}).
Os belongs to the third peak of the r-process elements and not many measurements can be found in the literature.
We derived the \ion{Os}{i} line at 442.0\,nm in 11 stars, and the [Os/Fe] ratio varies from 0.28 to 0.85\,dex (see Fig.\,\ref{fig:osfe}).
If we assume that Y is a pure s-process element and Eu a pure r-process element, the fact that
the abundance ratios of other n-capture elements show either no correlation with either
or a statistically significant correlation with both, suggests that both processes
are responsible for the nucleosynthesis of these elements in the metallicity range
explored by our sample of stars. The only exception is Sm, that displays a strong
correlation with [Eu/Fe] but none with [Y/Fe].
We take this as an indication that, contrary to what happens in the solar system, where Sm
is formed for about 70\% by the r-process and 30\% by the s-process \citep{arlandini99}, in this
metallicity range the Sm production is only due to the r-process.

\section*{Aknowledgements}

The authors wish to thank the referee for the useful comments.
PB acknowledges support   from the ERC advanced grant N. 835087 -- SPIAKID.
This research has made use of the SIMBAD database, operated at CDS, Strasbourg, France.

\bibliographystyle{aa}
\bibliography{biblio}

\section{Appendix: On-line tables\label{app}}

The detailed abundances are available in a csv table at CDS.

The atomic lines used in the analysis are presented in an ASCII table available at CDS.
Be careful, in the analysis performed by \mygi\ the observed spectrum is fitted in a selected region,
so that other lines can be present in the wavelength range.
The atomic data are from \citet{heiter21}.

\begin{table*}
\centering
\caption{Coordinates, Gaia DR3 ID and $G$ magnitudes, extinction A0, Gaia\,DR3 radial velocities with their uncerianties for the program stars.}
\label{coord}
\begin{tabular}{lrrlllrl}
\hline
  \multicolumn{1}{c}{Name} &
  \multicolumn{1}{c}{Gaia DR3 ID} &
  \multicolumn{1}{c}{$G$} &
  \multicolumn{1}{c}{RA (J2000)} &
  \multicolumn{1}{c}{DEC (J2000)} &
  \multicolumn{1}{c}{A0} &
  \multicolumn{1}{c}{$\rm V_r$ Gaia} &
  \multicolumn{1}{c}{$\rm \sigma(V_r)$} \\
 & & & & & & \kms & \kms \\
\hline
  RVS1383 & 4821430166509061632 & 11.555 & 05:25:23.71 & --37:15:54.56 & 0.054 & $ 540.36$ & 1.34 \\
  RVS1280 & 2966726811218049280 & 9.972 & 05:54:57.74 & --18:29:19.35  & 0.113 & $ 503.53$ & 0.36 \\
  RVS1392 & 2921543205513614208 & 12.081 & 06:59:45.23 & --24:26:31.98 & 0.093 & $ 591.28$ & 0.47 \\
  RVS1315 & 5717948445741886720 & 9.506 & 07:59:46.83 & --17:23:08.35  & 0.089 & $ 510.25$ & 0.29 \\
  RVS1358 & 5724454084243854464 & 11.675 & 07:59:50.63 & --15:38:05.33 & 0.106 & $ 515.70$ & 0.37 \\
  RVS924 & 5642811738112717824 & 14.063 & 08:45:25.26 & --28:37:16.49  & 0.214 & $ 587.07$ & 3.57 \\
  RVS920 & 5642140073946549120 & 13.970 & 08:46:47.12 & --29:46:24.09  & 0.307 & $ 537.88$ & 2.87 \\
  RVS1281 & 5325632011075073408 & 12.238 & 09:13:01.84 & --50 08 56.77 & 3.500 & $ 600.86$ & 0.61 \\
  RVS918 & 5218683236590712192 & 14.164 & 09:14:25.73 & --72:35:08.97  & 0.243 & $ 512.44$ & 2.45 \\
  RVS917 & 5430165914418381952 & 13.905 & 09:26:15.75 & --37:55:35.29  & 0.352 & $ 542.97$ & 2.33 \\
  RVS916 & 5662947644299064960 & 14.007 & 09:34:15.15 & --23:18:26.47  & 0.113 & $ 508.96$ & 2.13 \\
  RVS932 & 5415294332460034944 & 14.156 & 10:11:19.62 & --42:47:36.09  & 0.274 & $ 509.45$ & 1.54 \\
  RVS933 & 5358307297639362304 & 13.417 & 10:21:32.77 & --51:47:52.91  & 0.838 & $ 559.87$ & 1.63 \\
  RVS938 & 5447654677647394432 & 13.858 & 10:28:43.42 & --33:10:27.50  & 0.133 & $ 514.75$ & 3.44 \\
  RVS928 & 3778236623817788416 & 12.837 & 10:44:08.72 & --03:05:51.67  & 0.092 & $ 543.45$ & 2.39 \\
  RVS941 & 5397224473030069504 & 13.533 & 11:17:14.68 & --37:11:32.91  & 0.256 & $ 508.80$ & 2.01 \\
  RVS911 & 5372997906848376320 & 12.845 & 11:31:17.08 & --49:36:58.83  & 0.302 & $ 527.52$ & 0.55 \\
  RVS937 & 5382844170907456128 & 13.732 & 11:32:09.58 & --41:04:22.45  & 0.251 & $ 531.97$ & 1.20 \\
  RVS922 & 5370652682908661760 & 13.829 & 11:55:45.24 & --50:20:40.02  & 0.258 & $ 539.52$ & 2.31 \\
  RVS914 & 6073933580264025856 & 13.657 & 12:57:01.33 & --53:58:46.92  & 0.744 & $ 515.66$ & 2.76 \\
  RVS923 & 5791578032174490752 & 13.905 & 13:51:18.83 & --73:12:22.29  & 0.802 & $ 511.59$ & 1.45 \\
  RVS926 & 4169127503008440960 & 14.311 & 17:43:37.46 & --06:29:46.27  & 2.880 & $-513.06$ & 2.16 \\
  RVS919 & 6435715065193423360 & 13.775 & 18:00:54.95 & --68:58:24.40  & 0.224 & $ 510.63$ & 1.55 \\
  RVS931 & 4236913258154716032 & 13.702 & 20:02:07.41 & --00:20:38.40  & 0.570 & $-509.32$ & 1.43 \\
  RVS930 & 4246534637728074112 & 13.530 & 20:30:33.01 & +06:02:40.52   & 0.220 & $-559.14$ & 2.77 \\
\hline\end{tabular}
\end{table*}

\end{document}